\documentclass[runningheads]{llncs}
\usepackage{graphicx}
\usepackage{multirow}
\usepackage{tabularx}
\newcolumntype{Y}{>{\centering\arraybackslash}X}

\begin{document}
\title{Data augmentation for dealing with low sampling rates in NILM}
%
%
\author{Tai Le Quy \and
Sergej Zerr \and
Eirini Ntoutsi \and
Wolfgang Nejdl}
\authorrunning{T. Le Quy et al.}
%
\institute{L3S Research Center - Leibniz University Hannover, Germany\\
\email{\{tai, szerr, ntoutsi, nejdlg@l3s.de\}}
}
\maketitle              
\begin{abstract}
Data have an important role in evaluating the performance of NILM algorithms. The best performance of NILM algorithms is achieved with high-quality evaluation data. However, many existing real-world data sets come with a low sampling quality, and often with gaps, lacking data for some recording periods. As a result, in such data, NILM algorithms can hardly recognize devices and estimate their power consumption properly. An important step towards improving the performance of these energy disaggregation methods is to improve the quality of the data sets. In this paper, we carry out experiments using several methods to increase the sampling rate of low sampling rate data. Our results show that augmentation of low-frequency data can support the considered NILM algorithms in estimating appliances’ consumption with a higher F-score measurement.

\end{abstract}
\section{Introduction}
\label{sec:introduction}

Non-Intrusive Load Monitoring (NILM) or energy disaggregation is a state-of-the-art technology to disaggregate and estimate the power consumption of individual appliances from the aggregated signal in households or companies. It is preferred over intrusive approaches due to bounded costs compared to the monitoring of each device separately. From the first work of Hart~\cite{hart1992nonintrusive}, a number of NILM techniques have been proposed~\cite{zoha2012non}. NILM research requires large amounts of high-quality aggregated data~\cite{beckel2014eco,klemenjak2016non}. It has been shown that NILM algorithms work efficiently with a sampling rate at 1Hz because at this granularity the data can provide several electricity measurements such as active and reactive power~\cite{beckel2014eco,klemenjak2016non}. However, many data sets exist with lower sampling rates, like Pecan Street Sample~\cite{holcomb2012pecan} at 1/60 Hz or HES~\cite{zimmermann2012household} at 1/120 and 1/600 Hz. Recognizing devices in such data sets is extremely difficult for the existing NILM algorithms. Besides the low sampling rates, the algorithms have often to deal with data incompleteness like missing aggregated signal for some time points or even, for certain time periods. Furthermore, recently, deep learning techniques have shown their potential for NILM, e.g., ~\cite{kelly2015neural,mauch2015new}, however, they require huge amounts
of training data. As a result, constructing higher sampling rate data even from a small amount of the lower samples is an important direction in order to improve the performance of NILM algorithms.

In this research, we investigate and propose data augmentation methods in order to construct high sampling rate data from the low sampling rate signal that can be used for NILM. To this end, we apply interpolation techniques such as Denton-Cholette method for temporal disaggregation, stepwise method and Cubic spline interpolation on power consumption data of two selected data sets (ECO~\cite{beckel2014eco} and iAWE~\cite{batra2013s}) in order to generate a higher sampling rate data. We then report the results
of the performance of two NILM methods (Weiss’s algorithm~\cite{weiss2012leveraging} and Parson’s algorithm~\cite{parson2012non}). 

The rest of the paper is structured as follows: In Section~\ref{sec:relatedwork}, we overview the related work. In Section~\ref{sec:method}, we describe the augmentation techniques for generating high sampling rate data from the lower samples. In Section~\ref{sec:evaluation}, we present our experimental results as the effect of augmentation on the performance of two well-known NILM algorithms. Conclusions and outlook are finally presented in Section~\ref{sec:conclusion}.

\section{Related work}
\label{sec:relatedwork}

The first NILM method has been introduced by Hart~\cite{hart1992nonintrusive} to extract device consumption profiles called signatures. Following this work, different methods have been proposed relying on state analysis (e.g., tracking on/off operation by using real power and reactive power), utilize learning methods or different data granularities~\cite{klemenjak2016non,zoha2012non}. Parson et al.~\cite{parson2012non} proposed a semi-supervised approach using factorial HMMs which was
evaluated on data sets at a sampling rate 1/60 Hz. Weiss et al.~\cite{weiss2012leveraging} proposed a supervised approach to extract switching events and find the best match in a signature database by using
real and reactive power information with granularity 1 Hz. 

Unfortunately, the publicly available data sets do not always come with the desired granularity. In fact, data signal collection is a very costly process, in terms of the time required to collect reasonably large data sets but also due to other reasons like privacy. In Table~\ref{tbl:datasets}, we survey the data sets that are often used for NILM evaluation. As depicted, different datasets come with different granularities and for some cases, the sampling rate is too low. 
\begin{table*}[!htb]
\caption{Datasets for NILM}
\label{tbl:datasets}
\begin{center}

\begin{tabularx}{0.8\textwidth}{ |c| *{3}{Y|} }
\hline
Dataset &  Institution & Sampling rate\\
\hline
REDD (2011) &  MIT  & 1 Hz and 15 kHz\\
BLUED (2012) & CMU & 12 kHz \\
Smart* (2012) & UMass & 1 Hz \\
Sample (2013) & Pecan Street & 1/60 Hz \\
HES (2013) &  DECC DEFRA & 1/120 Hz and 1/600 Hz \\
iAWE (2013) & III Delhi & 1 Hz \\
ECO (2013) & ETH & 1 Hz \\
UK-DALE (2014) & Imperial College & 1/6 Hz and 16 kHz \\
\hline
\end{tabularx}
\end{center}
\end{table*}

In contrast to existing works focussing on better NILM methods to cope with real-world electricity data, we follow an approach by proposing data augmentation for NILM in order to generate high granularity samples from low granularity ones. Except for the low sample case, such an augmentation can also help with data incompleteness. Our method independent approach can work with a variety of different NILM algorithms because it is applied at the data level. 

Data augmentation is summarizing techniques for dealing with data sparsity by deducing missing values using historical information or third-party information sources. Recently, it has gained a lot of attention as a way to cope with the huge data demands of complex learning models such as deep neural
networks, and it has been proven that for many architectures and different applications, it improves the robustness and the generalization power of the models. Data augmentation techniques increase the volume of the data by generating new instances from the existing instances. In the image domain,
for example, this is achieved by applying label-preserving transformations like rotation and illumination~\cite{krizhevsky2012imagenet} to teach a machine model and achieve higher accuracy. Similarly, in the audio domain augmentation is achieved by applying transformations like adding background noise or changing the pitching level~\cite{salamon2017deep}. In the case of signal data being undersampled, augmentation is related to interpolation, which is used traditionally for e.g., time series and trajectory data~\cite{macedo2008trajectory}. In trajectory data, for example, interpolation between successive moving object positions is used in order to simulate the continuous nature of the actual movement.

To our best knowledge, data augmentation for NILM has not been investigated thus far, however, our results show that it comprises a promising approach.

\section{Generating high sampling rate data from low rate samples}
\label{sec:method}

In our work, the augmentation methods are used to generate augmented time series data in between every timestamp $t_{i}$ and $t_{i+1}$.

\subsection{Stepwise interpolation}
Augmented data in between every time-stamp ti and $t_{i+1}$ is generated by dividing the time-gap of $t_i$ and $t_{i+1}$ into $k$ parts, then we create data for each part by the following formula:
\begin{equation}
    data(part_j) = data(t_i) + \frac{data(t_{i+1})-data(t_i)}{k}*j ,j = 1..k
\end{equation}

\subsection{Cubic Spline interpolation}
This is a form of interpolation using a special type of piece-wise polynomial called a spline. In our work, we use a spline function in \textit{Numerical Recipes in C}\footnote{http://www.aip.de/groups/soe/local/numres/bookcpdf/c3-3.pdf}, that returns interpolated values between data at time-stamp $t_i$ and $t_{i+1}$. The distance to the next interpolated point is calculated by a parameter $t$ with range from 0 for time-stamp $t_i$ to 1 for timestamp $t_{i+1}$.

\subsection{Denton-Cholette interpolation}
This is a method for temporal disaggregation, it can disaggregate low-frequency time series data with or without high-frequency indicator series. This method is primarily concerned with movement preservation. Augmented data that is similar to the indicator is generated by considering the most common case in the indicator. Mathematical techniques are used to distribute low-frequency data in high-frequency series when the indicator has a poor quality. In this paper, we use an implementation of this method in a
R package by Christoph et al.~\cite{saxtemporal}.
\subsection{Device interpolation}
In this method, the changes in the power consumption of devices with high sampling rates are used as indicators to estimate the augmented values for aggregated signals. For every time points in between two time-stamps $t_i$ and $t_{i+1}$, if the power consumption of any appliance changes beyond a threshold value. In our experiment, we set the threshold value from 5 to 10W. We will increase or decrease the aggregated value to the power level of time-stamp $t_{i+1}$. Otherwise we set the same value for every time series data in between $t_i$ and $t_{i+1}$.

\begin{figure}[h]
\includegraphics[width=\textwidth]{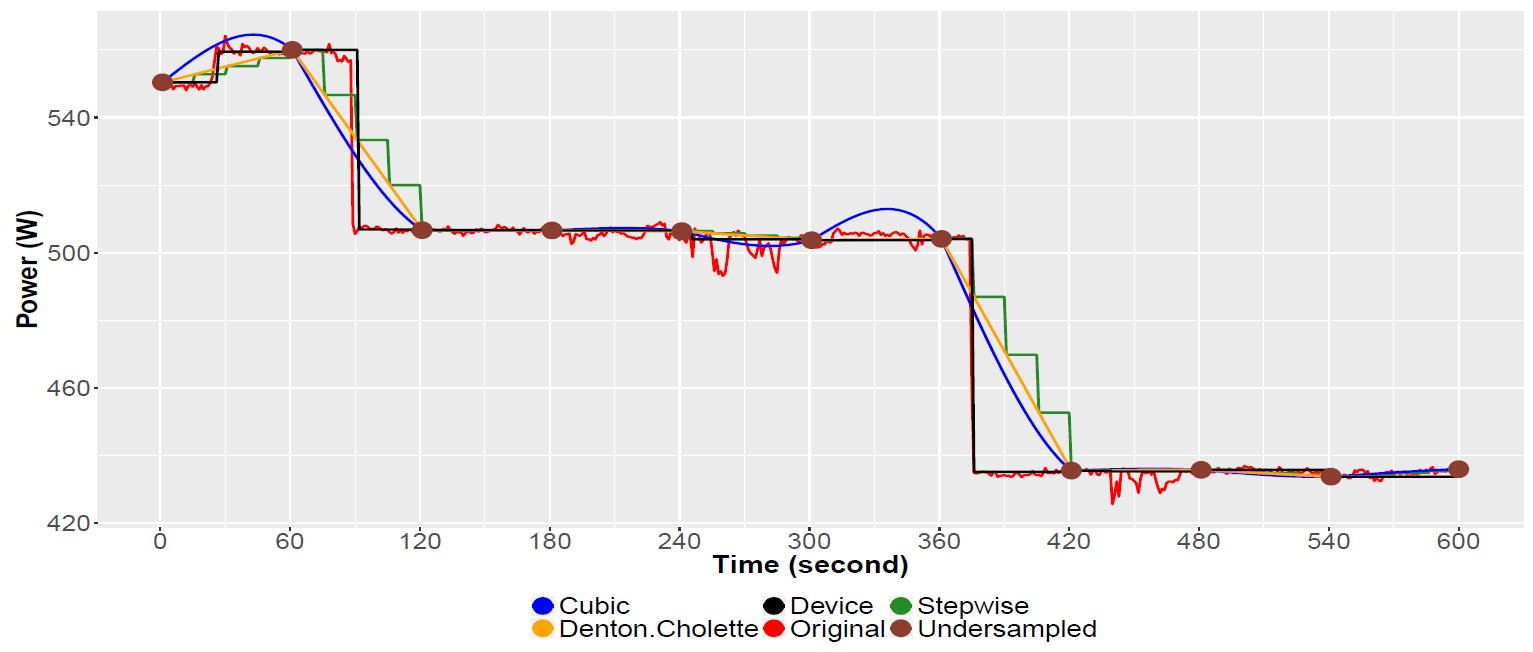}
\caption{A sample augmented data with different methods} 
\label{fig:sample}
\end{figure}

Fig.~\ref{fig:sample} visualizes 10 minutes of aggregated data on 01/06/2012 in house 2 of ECO data set with different augmentation methods.
\section{Experiments}
\label{sec:evaluation}
We evaluate the effect of data augmentation for NILM on two real data sets (Section~\ref{subsec:datasets}), the evaluation is presented in Section~\ref{subsec:result}.
\subsection{Datasets}
\label{subsec:datasets}
The \textbf{iAWE dataset}~\cite{batra2013s} contains ambient, water and electricity data from a single house in Delhi over a period of 73 days (May-August 2013), 10 jPlugs are used to measure 10 appliances with multiple parameters including voltage, current, phase and frequency. It consists of almost 15M measurements at a sampling rate 1 Hz. The average of missing values in this data set is 31\%.

The \textbf{ECO dataset}~\cite{beckel2014eco} contains more than 650M measurements from 45 smart plugs in 6 households in Switzerland over a period of 8 months (June 2012-January 2013). The data set is collected at a sampling rate 1 Hz with the aggregated consumption data and appliances’ consumption. Each of the measurements contains information on power consumption, voltage, current and the phase shift between voltage and current. For each household, there are 6-10 devices connected to smart plugs to measure power consumption in order to obtain ground truth data for analysis. The average of missing values in smart meter data is 0.8\%. The proportion of appliance consumption measured by plug meters in households varies from 10\% to 80\%.

These datasets contain both real and reactive power measurements which are a prerequisite for several NILM algorithms, e.g., Weiss et al.~\cite{weiss2012leveraging}. For \textbf{iAWE} dataset, we use the whole observation period (73 days). We evaluate the performance of NILM algorithms on five appliances: two air conditionals, fridge, laptop-PC and TV. In the \textbf{ECO} dataset, each house has a different observation period. Therefore, we select 30 days for each house as follows: House 1, 2: from 01/06/2012 to 30/06/2012, House 3: from 06/12/2012 to 04/01/2013, House 4, 5, 6: from 01/07/2012
to 30/07/2012. We evaluate the performance on 15 devices: dryer, freezer, fridge, water kettle, PC, laptop, dishwasher, lamp, microwave, stove, stereo system, TV, coffee machine, entertainment system and fountain. For evaluation, we took the first 2/3 of the recording period from each data set for training and used the remaining segment for testing. Resulting in 20 and 50 days training data for \textbf{ECO} and \textbf{iAWE} datasets respectively.

\subsection{Experimental setup}
\label{subsec:setup}
We evaluate the performance, in terms of the F-score on the estimation of power consumption, of two well-known NILM algorithms, namely Parson~\cite{parson2012non} and Weiss~\cite{weiss2012leveraging}. For the experiments, we used the \textit{NILM-Eval framework}~\cite{beckel2014eco}.

To evaluate the effect of data augmentation for creating high sampling data from low sampling ones, we first \textit{down sample} the original data sets to the 1/60 Hz granularity by keeping the first second data for each minute. For Parson algorithm, because this algorithm is designed to work with
data at a sampling rate 1/60 Hz, we down-sample the data to the granularity of 1/600 Hz (10 minutes) before evaluating the performance of this algorithm. We refer to the undersampled data set as “undersampled” and to the original data set as “original”. For the experiment of Weiss algorithm, we generate the
augmented data at granularity 1Hz from the “undersampled” data at 1/60Hz. Parson algorithms use the data at the sampling rate 1/60Hz which is reconstructed from the “undersampled” data at granularity 1/600Hz. We also do several experiments to find the best parameter for our augmentation methods.
In the \textit{stepwise} method, we carried out the experiments with different values of k (k = 2; 3; 4; 6; 10) and we found that with k = 4 our stepwise interpolation shows the best results across the available datasets. We then compare the performance of the NILM algorithms for the different data sets:
“original”, “undersampled” and several proposed augmented methods. Our goal is to investigate, how augmentation helps the undersampled data set to reach a performance close to the
original high sampling data set.

\subsection{Performance evaluation}
\label{subsec:result}
Table~\ref{tbl:mse_rmse} describes the Mean Squared Error (MSE) and Root Mean Squared Error (RMSE) that shows the difference between the augmented data generated by our augmentation methods and the original data on two data sets. We calculate the average values of MSE and RMSE of 6 households in the
ECO data set. These measurements are calculated on the data set at granularity 1Hz.

\begin{table*}[!htb]
\caption{MSE and RMSE of augmentation methods}
\label{tbl:mse_rmse}
\begin{center}
\begin{tabularx}{0.9\textwidth}{ |c| *{4}{Y|} }
\cline{2-5}
\hline
\multirow{2}{*}{Augmentation methods} & \multicolumn{2}{c|}{ECO dataset} & \multicolumn{2}{c|}{iAWE dataset}                    \\ \cline{2-5} 
                         & \multicolumn{1}{c|}{MSE} & \multicolumn{1}{c|}{RMSE} & \multicolumn{1}{c|}{MSE} & \multicolumn{1}{c|}{RMSE} \\
\hline
Stepwise    & 85,374.7      & 286.7                 & 137,600.8                & 370.9 \\
Cubic & 247,425.5  &  473.7 &  \textbf{100,464.6}  & \textbf{316.9} \\
Denton-Cholette & \textbf{34,921.3}  & \textbf{181.9} & 125,423.4 &  354.1 \\
Device & 61,021.8 & 238.6 & 330,394.8 & 574.8 \\
\hline
\end{tabularx}
\end{center}
\end{table*}
\textbf{Comparison of the NILM Algorithm performance on original data}: During our experiments, we noticed the difference in performance and properties of the used NILM techniques. Whilst Parson algorithm was showing good performance for smaller period data sets, it was lagging behind for data sets comprising large time periods. The possible explanation is that Parson uses pre-trained models, whilst Weiss needs more time to train, but in the long run, it is able to identify more devices, especially outperforming at the “outlier-devices” such as Fountain and Dryer. For similar reasons, the smaller \textbf{iAWE} dataset was dominated by Parson and the larger \textbf{ECO} dataset by Weiss. In this context, while Parson showed high precision for a small number of devices, Weiss was able to identify more devices with moderate precision.

\begin{figure}[!htb]
\includegraphics[width=\textwidth]{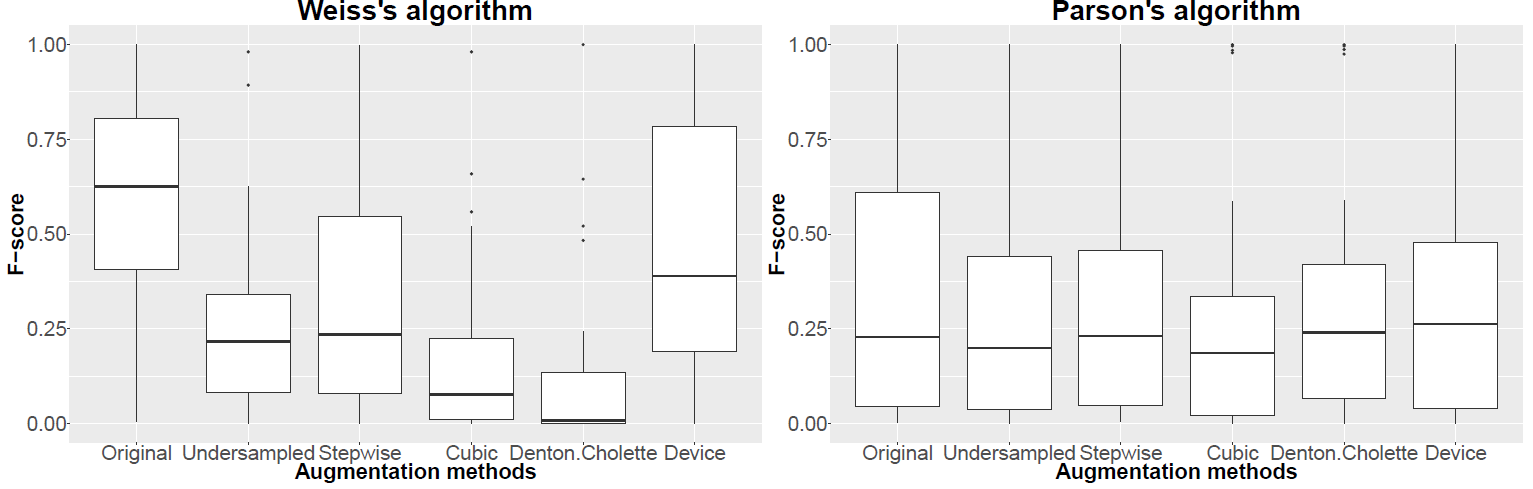}
\caption{Performance of NILM algorithms on ECO dataset} 
\label{fig:ECO_result}
\end{figure}

\textbf{Down-sampling effect}: Separately for each method, we measured the effect of the down-sampling. Down-sampling has shown a strong effect on the Weiss method, while it had a light effect on Parson algorithm. This can be explained by the fact that the Weiss method requires information about active power and reactive power as shown for the ECO data set in Figure~\ref{fig:ECO_result}

\begin{figure}[!htb]
\includegraphics[width=\textwidth]{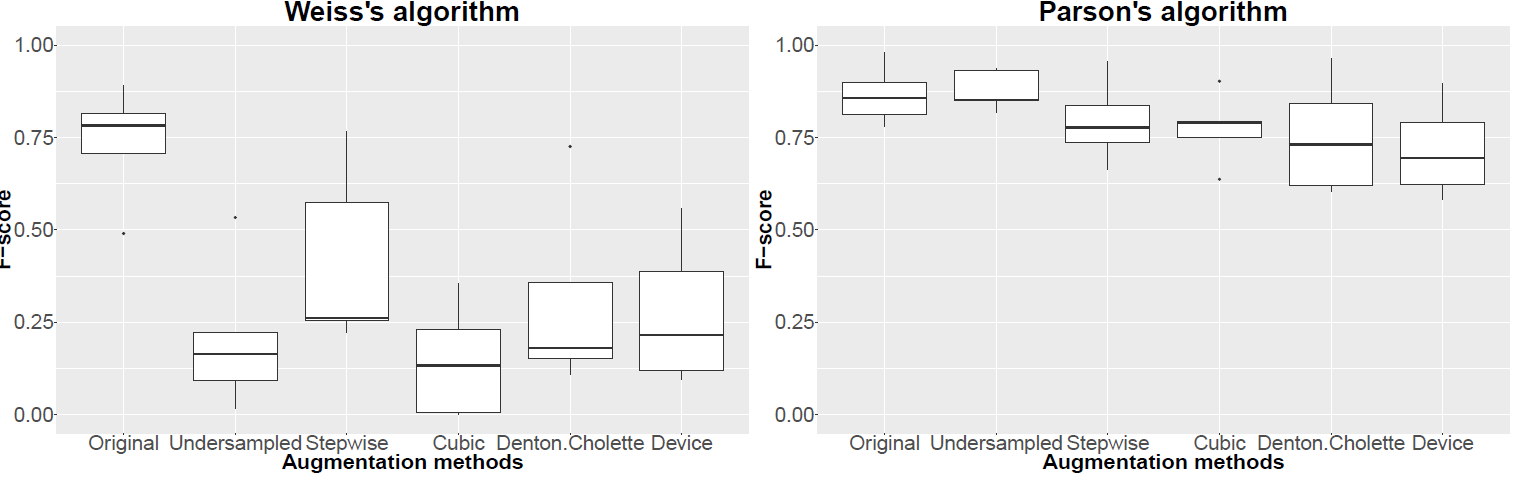}
\caption{Performance of NILM algorithms on iAWE dataset} 
\label{fig:iAWE_result}
\end{figure}

\textbf{Comparison of the augmentation methods}: The performance of the augmentation methods (measured by F-score
value) on iAWE data set are shown in Table~\ref{tbl:iawe_method} and Figure~\ref{fig:iAWE_result}
3, the results on ECO data set are presented in Table~\ref{tbl:ECO_method} and Figure~\ref{fig:ECO_result}. Comparing the different augmentation techniques, device interpolation shows the best performance for most of
the devices in both the Weiss and Parson algorithm. The stepwise method follows, whilst the cubic interpolation shows the worse performance among the remaining augmentation techniques. One of the reasons, as we noticed from the augmented signal results, is the introduced smoothness of the produced time series, which can hinder event detection and disturb the inferring power consumption of the appliances.

\begin{table*}[!htb]
\caption{Performance of NILM on augmentation methods on iAWE dataset}
\label{tbl:iawe_method}
\begin{center}
\begin{tabularx}{0.9\textwidth}{ |c| *{4}{Y|} }
\cline{2-5}
\hline
\multirow{2}{*}{Augmentation methods} & \multicolumn{2}{c|}{Weiss’s algorithm}                     & \multicolumn{2}{c|}{Parson’s algorithm}                    \\ \cline{2-5} 
                         & \multicolumn{1}{c|}{Avg.} & \multicolumn{1}{c|}{Dev.} & \multicolumn{1}{c|}{Avg.} & \multicolumn{1}{c|}{Dev.} \\
\hline
Original & 0.736 & 0.136 & 0.866 & 0.007 \\
Undersampled & 0.206 & 0.178 & 0.877 & 0.048 \\
Stepwise & \textbf{0.416} & 0.217 & \textbf{0.793} & 0.099 \\
Cubic & 0.145 & 0.136 & 0.775 & 0.085 \\
Denton-Cholette & 0.305 & 0.227 & 0.752 & 0.136 \\
Device & 0.275 & 0.175 & 0.718 & 0.114 \\
\hline
\end{tabularx}
\end{center}
\end{table*}

\begin{table*}[!htb]
\caption{Performance of NILM on augmentation methods on ECO dataset}
\label{tbl:ECO_method}
\begin{center}
\begin{tabularx}{0.9\textwidth}{ |c| *{4}{Y|} }
\cline{2-5}
\hline
\multirow{2}{*}{Augmentation methods} & \multicolumn{2}{c|}{Weiss’s algorithm}                     & \multicolumn{2}{c|}{Parson’s algorithm}                    \\ \cline{2-5} 
                         & \multicolumn{1}{c|}{Avg.} & \multicolumn{1}{c|}{Dev.} & \multicolumn{1}{c|}{Avg.} & \multicolumn{1}{c|}{Dev.} \\
\hline
Original & 0.604 & 0.257 & 0.359 & 0.331\\
Undersampled & 0.269 & 0.240 & 0.27 &  0.274 \\
Stepwise & 0.332 & 0.278 & 0.328 & 0.307 \\
Cubic & 0.176 & 0.236 & 0.277 & 0.323 \\
Denton-Cholette & 0.128 & 0.234 & 0.325 & 0.305 \\
Device & \textbf{0.467} & 0.331 & \textbf{0.331} & 0.310 \\
\hline
\end{tabularx}
\end{center}
\end{table*}

\textbf{Performance with respect to device type}: We categorize devices into five groups based on their function and characteristics: Cooling devices (Freezer, Fridge), Cooking devices (Coffee Machine, Microwave, Water kettle, Stove), Entertainment (TV, Stereo), Computer (PC, Laptop), Lighting device (Lamp). A summary of results for groups of appliances in ECO data set is presented in Table~\ref{tbl:ECO_appliance_weiss} and Table~\ref{tbl:ECO_appliance_parson}. The \textit{device} interpolation method showed the best performance among the four methods, although in some cases it failed for Lighting devices, as such appliances consume a low amount of electric power. Another observation is that Parson’s algorithm can work well with data generated by Device interpolation and Stepwise methods for cooling devices, entertainment and computer, as these appliances do not have significant changes in power consumption over time.

\begin{table*}[!htb]
\caption{Performance of Weiss’s alg. with appliance’s groups ECO dataset:}
\label{tbl:ECO_appliance_weiss}
\begin{center}
\begin{tabularx}{1\textwidth}{ |c| *{6}{Y|} }
\hline
Augmentation methods & Cooling & Cooking & Ent. & Computer & Lighting \\
\hline
Original & 0.711 & 0.613 & 0.585 & 0.357 & 0.281 \\
Undersampled & 0.318 & 0.101 & 0.396 & 0.208 & 0.174 \\
Stepwise & 0.482 & 0.231 & 0.312 & \textbf{0.045} & 0.157 \\
Cubic & 0.273 & 0.124 & 0.246 & 0.021 & 0.078 \\
Denton-Cholette & 0.239 & 0.037 & 0.181 & 0.014 & 0.060 \\
Device & \textbf{0.744} & \textbf{0.300} & \textbf{0.416} & 0.043 & \textbf{0.193} \\
\hline
\end{tabularx}
\end{center}
\end{table*}

\begin{table*}[!htb]
\caption{Performance of Parson’s alg. with appliance’s groups on ECO dataset}
\label{tbl:ECO_appliance_parson}
\begin{center}
\begin{tabularx}{1\textwidth}{ |c| *{6}{Y|} }
\hline
Augmentation methods & Cooling & Cooking & Ent. & Computer & Lighting \\
\hline
Original & 0.667 & 0.042 & 0.410 & 0.454 & 0.03 \\
Undersampled & 0.504 & 0.021 & 0.326 & 0.306 & 0.028 \\
Stepwise & 0.596 & 0.048 & 0.306 & \textbf{0.485} & \textbf{0.037} \\
Cubic & 0.482 & 0.038 & \textbf{0.319} & 0.373 & 0.023\\
Denton-Cholette & 0.574 & 0.047 & \textbf{0.319} & 0.472 & 0.036\\
Device & \textbf{0.620} & \textbf{0.049} & 0.309 & 0.456 & \textbf{0.037}\\
\hline
\end{tabularx}
\end{center}
\end{table*}

\section{Conlusion and outlook}
\label{sec:conclusion}
In this work, we presented an attempt to assist NILM by improving the quality of low sampling rate energy consumption data sets through data augmentation by using several interpolation techniques. Our approach works at the data level and therefore it is method-independent and applicable for a variety of different NILM algorithms. Augmentation was also shown to be helpful for data-intensive machine learning models like deep neural networks which recently have been successfully used also for NILM~\cite{kelly2015neural,mauch2015new}. Our results show that data augmentation is applicable for increasing the sampling rates of a data set. We believe that this is a promising direction for NILM and further research should be carried, in parallel to the development of new, more sophisticated methods.

In this preliminary work, we adapted simple augmentation techniques, which nevertheless yield improvements over the
non-augmented low-sample data sets. As part of our ongoing work, we are investigating more sophisticated augmentation
approaches which take into account the consumption profile of the household as well as the profiles of individual devices and do not require the high sampling rate data of appliances. Deep learning technology is also a potential approach that can learn from devices’ using patterns in order to construct the aggregated data. Such “informed”-augmentation approaches are expected to yield better augmented data and therefore, better predictions.

\section*{Acknowledgements}
The work of the first author is financially supported by the Ministry of Education and Training, Vietnam, within the program ``Training doctoral degrees for lecturers at universities and colleges from 2010 to 2020" (Project 911).
%
%
%
\bibliographystyle{splncs04}
\bibliography{mybibliography}

\begin{thebibliography}{10}
\providecommand{\url}[1]{\texttt{#1}}
\providecommand{\urlprefix}{URL }
\providecommand{\doi}[1]{https://doi.org/#1}

\bibitem{batra2013s}
Batra, N., Gulati, M., Singh, A., Srivastava, M.B.: It's different: Insights
  into home energy consumption in india. In: Proceedings of the 5th ACM
  Workshop on Embedded Systems For Energy-Efficient Buildings. pp.~1--8 (2013)

\bibitem{beckel2014eco}
Beckel, C., Kleiminger, W., Cicchetti, R., Staake, T., Santini, S.: The eco
  data set and the performance of non-intrusive load monitoring algorithms. In:
  Proceedings of the 1st ACM conference on embedded systems for
  energy-efficient buildings. pp. 80--89 (2014)

\bibitem{hart1992nonintrusive}
Hart, G.W.: Nonintrusive appliance load monitoring. Proceedings of the IEEE
  \textbf{80}(12),  1870--1891 (1992)

\bibitem{holcomb2012pecan}
Holcomb, C.: Pecan street inc.: A test-bed for nilm. In: International Workshop
  on Non-Intrusive Load Monitoring, Pittsburgh, PA, USA (2012)

\bibitem{kelly2015neural}
Kelly, J., Knottenbelt, W.: Neural nilm: Deep neural networks applied to energy
  disaggregation. In: Proceedings of the 2nd ACM international conference on
  embedded systems for energy-efficient built environments. pp. 55--64 (2015)

\bibitem{klemenjak2016non}
Klemenjak, C., Goldsborough, P.: Non-intrusive load monitoring: A review and
  outlook. Informatik 2016  (2016)

\bibitem{krizhevsky2012imagenet}
Krizhevsky, A., Sutskever, I., Hinton, G.E.: Imagenet classification with deep
  convolutional neural networks. Advances in neural information processing
  systems  \textbf{25},  1097--1105 (2012)

\bibitem{macedo2008trajectory}
Macedo, J., Vangenot, C., Othman, W., Pelekis, N., Frentzos, E., Kuijpers, B.,
  Ntoutsi, I., Spaccapietra, S., Theodoridis, Y.: Trajectory data models. In:
  Mobility, Data Mining and Privacy, pp. 123--150. Springer (2008)

\bibitem{mauch2015new}
Mauch, L., Yang, B.: A new approach for supervised power disaggregation by
  using a deep recurrent lstm network. In: 2015 IEEE Global Conference on
  Signal and Information Processing (GlobalSIP). pp. 63--67. IEEE (2015)

\bibitem{parson2012non}
Parson, O., Ghosh, S., Weal, M., Rogers, A.: Non-intrusive load monitoring
  using prior models of general appliance types. In: Proceedings of the AAAI
  Conference on Artificial Intelligence. vol.~26 (2012)

\bibitem{salamon2017deep}
Salamon, J., Bello, J.P.: Deep convolutional neural networks and data
  augmentation for environmental sound classification. IEEE Signal Processing
  Letters  \textbf{24}(3),  279--283 (2017)

\bibitem{saxtemporal}
Sax, C., Steiner, P.: Temporal disaggregation of time series. A peer-reviewed,
  open-access publication of the R Foundation for Statistical Computing p.~80

\bibitem{weiss2012leveraging}
Weiss, M., Helfenstein, A., Mattern, F., Staake, T.: Leveraging smart meter
  data to recognize home appliances. In: 2012 IEEE International Conference on
  Pervasive Computing and Communications. pp. 190--197. IEEE (2012)

\bibitem{zimmermann2012household}
Zimmermann, J.P., Evans, M., Griggs, J., King, N., Harding, L., Roberts, P.,
  Evans, C.: Household electricity survey: A study of domestic electrical
  product usage. Intertek Testing \& Certification Ltd pp. 213--214 (2012)

\bibitem{zoha2012non}
Zoha, A., Gluhak, A., Imran, M.A., Rajasegarar, S.: Non-intrusive load
  monitoring approaches for disaggregated energy sensing: A survey. Sensors
  \textbf{12}(12),  16838--16866 (2012)

\end{thebibliography}


@article{hart1992nonintrusive,
  title={Nonintrusive appliance load monitoring},
  author={Hart, George William},
  journal={Proceedings of the IEEE},
  volume={80},
  number={12},
  pages={1870--1891},
  year={1992},
  publisher={IEEE}
}
@article{zoha2012non,
  title={Non-intrusive load monitoring approaches for disaggregated energy sensing: A survey},
  author={Zoha, Ahmed and Gluhak, Alexander and Imran, Muhammad Ali and Rajasegarar, Sutharshan},
  journal={Sensors},
  volume={12},
  number={12},
  pages={16838--16866},
  year={2012},
  publisher={Multidisciplinary Digital Publishing Institute}
}
@inproceedings{beckel2014eco,
  title={The ECO data set and the performance of non-intrusive load monitoring algorithms},
  author={Beckel, Christian and Kleiminger, Wilhelm and Cicchetti, Romano and Staake, Thorsten and Santini, Silvia},
  booktitle={Proceedings of the 1st ACM conference on embedded systems for energy-efficient buildings},
  pages={80--89},
  year={2014}
}
@article{klemenjak2016non,
  title={Non-intrusive load monitoring: A review and outlook},
  author={Klemenjak, Christoph and Goldsborough, Peter},
  journal={Informatik 2016},
  year={2016},
  publisher={Gesellschaft f{\"u}r Informatik eV}
}
@inproceedings{holcomb2012pecan,
  title={Pecan street inc.: A test-bed for nilm},
  author={Holcomb, Chris},
  booktitle={International Workshop on Non-Intrusive Load Monitoring, Pittsburgh, PA, USA},
  year={2012}
}
@article{zimmermann2012household,
  title={Household Electricity Survey: A study of domestic electrical product usage},
  author={Zimmermann, Jean-Paul and Evans, Matt and Griggs, Jonathan and King, Nicola and Harding, Les and Roberts, Penelope and Evans, Chris},
  journal={Intertek Testing \& Certification Ltd},
  pages={213--214},
  year={2012}
}
@inproceedings{kelly2015neural,
  title={Neural nilm: Deep neural networks applied to energy disaggregation},
  author={Kelly, Jack and Knottenbelt, William},
  booktitle={Proceedings of the 2nd ACM international conference on embedded systems for energy-efficient built environments},
  pages={55--64},
  year={2015}
}
@inproceedings{mauch2015new,
  title={A new approach for supervised power disaggregation by using a deep recurrent LSTM network},
  author={Mauch, Lukas and Yang, Bin},
  booktitle={2015 IEEE Global Conference on Signal and Information Processing (GlobalSIP)},
  pages={63--67},
  year={2015},
  organization={IEEE}
}
@inproceedings{batra2013s,
  title={It's Different: Insights into home energy consumption in India},
  author={Batra, Nipun and Gulati, Manoj and Singh, Amarjeet and Srivastava, Mani B},
  booktitle={Proceedings of the 5th ACM Workshop on Embedded Systems For Energy-Efficient Buildings},
  pages={1--8},
  year={2013}
}
@inproceedings{weiss2012leveraging,
  title={Leveraging smart meter data to recognize home appliances},
  author={Weiss, Markus and Helfenstein, Adrian and Mattern, Friedemann and Staake, Thorsten},
  booktitle={2012 IEEE International Conference on Pervasive Computing and Communications},
  pages={190--197},
  year={2012},
  organization={IEEE}
}
@inproceedings{parson2012non,
  title={Non-intrusive load monitoring using prior models of general appliance types},
  author={Parson, Oliver and Ghosh, Siddhartha and Weal, Mark and Rogers, Alex},
  booktitle={Proceedings of the AAAI Conference on Artificial Intelligence},
  volume={26},
  number={1},
  year={2012}
}
@article{krizhevsky2012imagenet,
  title={Imagenet classification with deep convolutional neural networks},
  author={Krizhevsky, Alex and Sutskever, Ilya and Hinton, Geoffrey E},
  journal={Advances in neural information processing systems},
  volume={25},
  pages={1097--1105},
  year={2012}
}
@article{salamon2017deep,
  title={Deep convolutional neural networks and data augmentation for environmental sound classification},
  author={Salamon, Justin and Bello, Juan Pablo},
  journal={IEEE Signal Processing Letters},
  volume={24},
  number={3},
  pages={279--283},
  year={2017},
  publisher={IEEE}
}
@incollection{macedo2008trajectory,
  title={Trajectory data models},
  author={Macedo, J and Vangenot, Christelle and Othman, Walied and Pelekis, Nikos and Frentzos, Elias and Kuijpers, Bart and Ntoutsi, Irene and Spaccapietra, Stefano and Theodoridis, Yannis},
  booktitle={Mobility, Data Mining and Privacy},
  pages={123--150},
  year={2008},
  publisher={Springer}
}
@article{saxtemporal,
  title={Temporal Disaggregation of Time Series},
  author={Sax, Christoph and Steiner, Peter},
  journal={A peer-reviewed, open-access publication of the R Foundation for Statistical Computing},
  pages={80}
}

\end{document}